\documentstyle [12pt]{article}
\newcommand{\cs}[3]{{{#3} \brace {#1 #2}}}

\newcommand{\h}[1]{\mathop{\lambda}\limits_{#1}\ \!\!\!}

\newcommand{\edf}{\ {\mathop{=}\limits^{\rm def}}\ }

\begin{document}
\title {The Spinning Equations of Motion for Objects  in AP-Geometry}
\maketitle
\begin{center}
{\bf {Magd E. Kahil{\footnote{,  Modern Sciences and Arts University, Giza, Egypt \\ Nile University, Giza, Egypt \\Egyptian Relativity Group, Cairo, Egypt \\
e.mail: mkahil@msa.eun.eg}}} }
\end{center}

\begin{abstract}
Equations of spinning objects are obtained in Absolute Parallelism Geometry [AP], a special class of non-Riemannian geometry admitting an  alternative non-vanishing curvature and torsion simultaneously. This new set of equations is the counterpart of the Papapetrou equations in the Riemannian geometry. Applying, the concept of geometerization of physics, it may give rise to describe the spin tensor as parameterized commutation relation between path and path deviation equations in both Riemannian and non-Riemannian geometries.

\end{abstract}
\section{ Introduction}
The problem of a rotating object in the presence of the gravitational field is essentially practical than viewing objects as a mere test particles, in order to ignore their intrinsic property due to the Orthodox General Theory of Relativity. Accordingly, several attempts were done  in the last century started by Mathisson [1], followed by Papapetrou [2] and extended by Dixon [3] to include other non-gravitational fields e.g. electromagnetic. Also, there is an approach by Dixon-Souriau to include spinning motion, magnetic moment with charged objects [4].
Such of these detailed equations have been presented only in Riemannin geometry.

Now, the arising question, is based on the following: \\
What is the situation of the above mentioned particles in case of Non-Remannain geometries?\\
 In order to find the above enquiry, one must take treat the situation of non-Riemanian geometries as individual cases:  one of its special classes is  Riemann-Cartan geometry; which considers a tetrad space $\h{i}^{\mu}$ as two independent vector fields, one may be responsible for general coordinate transformation (GCT), the holonomic coordinates, labeled by Greek indices and  the Latin ones are used to express the Local Lorentz transformation [LLT], mainly to describe the internal properties of the object [5], labeled by Latin letters, the anholonomic coordinates.  This type of work has led many authors [6-8], to relate this type of geometry with gauge theories of gravity [9]. In there are a tetrad space for gauge translation, and spin connection to represent gauge rotation. [10-12][..].\\

 Also, another trend of viewing the Non-Riemaniann geometry, is called a Teleparallel geometry- A geometry with a tetrad building blocks, may represent a translational gauge with a vanishing curvature [13] , and treating the annholonomic coordinates as vector number. Such a tendency of neutralizing the role of annholnomic coordinates, to be a vector number, with an additional property. This may give rise to define that there are non-vanishing torsion curvatures  simultaneously, due to within different types of absolute derivatives\footnote{For more details  about the underlying geometry and its application in establishing a generalized field theory see [14-16]}.

 The arising notation of AP-geometry led Wanas et al (1995) to describe three different paths may act the role of geodesic in Riemanian geometry [17]. The striking features of these paths, have a step $\frac{1}{2}$ from one path into other . This gives an impression, that paths in this type of geometry are naturally quantized. Lately, Wanas(1998) obtained a parameterized absolute parallelism geometry [PAP] obtaining  a spin-torsion interaction, together with defining non-vanishing curvature and torsion tensors simultaneously [18]. The existence of such an interaction has  led Wanas et al to detect its presence in terms of revealing the discrepancy between theory and observation of thermal neutrons [19] and presenting a temporal model for SN1987A [20].

 Accordingly, in the present work we are going to obtain the analogous of the Papapetrou equation with precesession in the context of AP-geometry.
 This will enable us to examine, the effect of different absolute derivatives on the interaction with the torsion and spin tensors. \\
The paper is organized as follows; section 2 displays the relationship between spin tensor and geodesic and geodesic deviation vectors in the Riemaniann geometry, section 3 is extending the previous relationship to become among paths and path deviation vectors with their corresponding spin tensors in AP-geometry. Section 4 deals with the Lagranagian formalism of the Papapetrou equation in AP-geometry, and finally, Section 5 displays the results obtained the previous sections , regarding some recommendations   in  our  future work of on this approach.

\section{Motion in Riemanian Geometry}

\subsection{ Geodesic and Geodesic Deviations : The Bazanski Approach}
 Equations of geodesic and geodesic deviation equations Riemannian geometry are required to examine many problems of motion for different test particles in gravitational fields. This led many authors to derive them by various methods, one of the most applicable ones is the Bazanski approach [21]  in which from one single Lagrangian one can obtain simultaneously equation of geodesic and geodesic deviations in the following way;
 \begin{equation}
 L = g_{\mu\nu} U^{\mu} \frac{D \Psi^{\nu}}{D s}
 \end{equation}
where , $g_{\mu \nu}$ is the metric tensor, $U^{\mu}$, is a unit tangent vector of the path whose parameter is $s$, and $\Psi^{\nu}$ is the deviation vector associated to the path $(s)$, $ \frac{D}{Ds}$ is the covariant derivative with respect to parameter $s$.
Applying the Euler Lagrange equation , by taking the variation with respect to the deviation tensor:

\begin{equation}
\frac{d}{ds} \frac{\partial L}{\partial \dot{\Psi}^{\mu}}- \frac{\partial L}{\partial {\Psi}^{\mu}} =0
 \end{equation}
to obtain the geodesic equation
\begin{equation}
\frac{D U^{\mu}}{D s} = 0
 \end{equation}
and taking the variation with respect the the unit vector $U^{\mu}$
\begin{equation}
\frac{d}{ds} \frac{\partial L}{\partial{U}^{\mu}}- \frac{\partial L}{\partial {x}^{\mu}} =0
 \end{equation}
to obtain the geodesic deviation equation
\begin{equation}
\frac{D^2 {\Psi}^{\mu}}{D s^2} = R^{\mu}_{\nu \rho \sigma} U^{\nu}U^{\rho}\Psi^{\sigma}
 \end{equation}
where $ R^{\mu}_{\nu \rho \sigma}$ is Riemann-Christoffel tensor.
\subsection{On The Relation Between Spin Tensor and The  Deviation Vector: The Riemanian Case}
Equations of spinning motion ,the case of $P^{\alpha}=mU^{\alpha}$ can be related to geodesic if one follows the following transformation [22]
\begin{equation}
\bar{U}^{\mu} = U^{\mu} + \beta \frac{D \Psi^{\mu}}{Ds}
 \end{equation}
where $\bar{U}^{\alpha}$ is  a unit tangent vector with respect to the parameter ,such that  $\bar{U}^{\alpha} = \frac{d x^{\mu}}{d \bar{s}} $ , $\bar{s}$.
By taking the covariant derivative on both sides one obtains:
\begin{equation}
\frac{D \bar{U}^{\alpha}}{D \bar{s}}= \frac{D}{Ds}(U^{\mu} + \beta \frac{D \Psi^{\mu}}{Ds})\frac{ds}{d \bar{s}}
 \end{equation}
From geodesic and geodesic deviation equations one gets
\begin{equation}
\frac{D U^{\alpha}}{Ds} =0
 \end{equation}
and
\begin{equation}
\frac{D^{2} \Psi^{\alpha}}{Ds^{2}} =R^{\alpha}_{\mu \nu \sigma} U^{\mu}U^{\nu}\Psi^{\sigma}
\end{equation}
Substituting equations (2.8) and (2.9) in  (2.)7 to get
\begin{equation}
\frac{D \bar{U}^{\alpha}}{D \bar{s}}=( \beta R^{\alpha}_{\mu \nu \sigma}U^{\mu} U^{\nu} \Psi^{\sigma})\frac{ds}{d \bar{s}}
 \end{equation}
Let us assume the following
taking $\beta =  \frac{s}{m}$
\begin{equation}
S^{\mu \nu} = s {(U^{\alpha }\Psi^{\beta}-U^{\beta }\Psi^{\alpha}) }
\end{equation}
Thus, we get
\begin{equation}
\frac{D \bar{U}^{\alpha}}{D \bar{s}}=\frac{1}{2m}( R^{\alpha}_{\mu \nu \sigma}U^{\mu} U^{\nu} \Psi^{\sigma})\frac{ds}{d \bar{s}}
 \end{equation}
i.e
\begin{equation}
\frac{D \bar{U}^{\alpha}}{D \bar{s}}= \frac{1}{2m} R^{\alpha}_{\mu \nu \sigma} S^{\nu \sigma} \bar{U}^{\mu}
 \end{equation}
Which is the Papapetrou equation for short.

\subsection{Lagrangian Formalism of Spinning Equations}
Another way to derive the Papapetrou equation for short is,  by applying the action principle  on the following equation [23]:
 \begin{equation}
 L = g_{\mu\nu} \bar{U}^{\mu} \frac{D \bar{\Psi}^{\nu}}{D \bar{s}} + \frac{1}{2m} R_{\mu \nu \rho \sigma} S^{\rho \sigma} U^{\nu}\Psi^{\mu},
 \end{equation}
taking the variation with respect to the deviation tensor $\bar{\Psi}^{\alpha}$ we obtain equation (2.13).

 Also, by taking the variation with respect to $\bar{U}^{\alpha}$ after some manipulations, we get its corresponding spinning deviation equation

 \begin{equation}
\frac{ D \bar{\Psi^2}}{D^2\bar{s}} = R^{\alpha}_{\beta \gamma \delta} \bar{U}^{\beta}\bar{U}^{\gamma}\bar{\Psi}^{\delta}+ \frac{1}{2m}(R^{\alpha}_{\beta \gamma \sigma} S^{\gamma \sigma}U^{\beta})_{; \delta}\bar{\Psi}^{\delta}
\end{equation}

Thus, we can figure out the Euler Lagrange Equations on the Bazanski-like Lagragian give an identical equation to (2.13)  and its corresponding deviation equations.

\subsection{Spinning and Spinning Deviation Equations Without Precession}
 The Papapetrou  equation of a spinning object with precession [2]is obtained  by a modified Bazanski Lagrangian [23] :
$$
L= g_{\alpha \beta} ( m U^{\alpha} + U_{\beta}\frac{D S^{\alpha \beta}}{Ds}) \frac{D \Psi^{\beta}}{Ds} + \frac{1}{2} R_{\alpha \beta \gamma \delta} S^{\gamma \delta} U^{\beta} \Psi^{\alpha}
$$
to obtain equation of a spinning object by taking the variation with respect to the deviation vector $\Psi^{\alpha}$
\begin{equation}
 \frac{D}{Ds}( m U^{\alpha} + U_{\beta}\frac{D S^{\alpha \beta}}{Ds})= \frac{1}{2} R^{\alpha}_{. \mu \nu \rho} S^{\rho \nu} U^{\mu}
\end{equation}
and its deviation equation can be obtained by taking the variation with respect to $U^{\alpha}$ to become:
$$
\frac{D^{2}\Psi^{\alpha}}{Ds^{2}}=  R^{\alpha}_{.\mu
\nu\rho}U^{\mu}( m U^{\nu} + U_{\beta}\frac{D S^{\nu
\beta}}{Ds})\Psi^{\rho}+ g^{\alpha \sigma}g_{\nu \lambda}( m
U^{\lambda} + U_{\beta}\frac{D S^{\lambda \beta}}{Ds})_{; \sigma}
\frac{D \Psi^{\nu}}{Ds}
$$
\begin{equation}
~~~~~~~+ \frac{1}{2}(R^{\alpha}_{. \mu \nu \rho} S^{\nu \rho} \frac{D
\Psi^{\mu}}{Ds}+ R^{\alpha}_{\mu \nu \lambda}S^{\nu \lambda}_{.;
\rho}U^{\mu}\Psi^{\rho} + R^{\alpha}_{\mu \nu \lambda; \rho
}S^{\nu \lambda} U^{\mu} \Psi^{\rho}) .
\end{equation}
\subsection{Spinning and Spinning Deviation Equations with Precession}
  It is well known that equation of  spinning charged  objects in the presence of gravitational field have been studied extensively [23]. This led us to suggest its corresponding  Lagrangian formalism , using a modified Bazanski Lagrangian [25], for  a spinning and precessing object and their corresponding deviation equation in Riemanian geometry in the following way
\begin{equation}
 L= g_{\alpha \beta} P^{\alpha} \frac{D \Psi^{\beta}}{Ds} + S_{\alpha \beta}\ \frac{D \Psi^{\alpha \beta}}{Ds}+ F_{\alpha}\Psi^{\alpha}+ M_{\alpha \beta}\Psi^{\alpha \beta}
 \end{equation}
 where
\begin{equation}
P^{\alpha}= m U^{\alpha}+ U_{\beta} \frac{D S^{\alpha \beta}}{DS}.
 \end{equation}
 Taking the variation with respect to $ \Psi^{\mu}$ and $\Psi^{\mu \nu}$ simultaneously we obtain
 \begin{equation}
\frac{DP^{\mu}}{DS}= F^{\mu},
 \end{equation}
and\begin{equation}
\frac{DS^{\mu \nu}}{DS}= M^{\mu \nu} ,
\end{equation}
 where $P^{\mu}$ is the momentum vector, $$ F^{\mu} = \frac{1}{2} R^{\mu}_{\nu \rho \delta} S^{\rho \delta} U^{\nu},$$ and $R^{\alpha}_{\beta \rho \sigma}$ is the Riemann curvature, $\frac{D}{Ds}$ is the covariant derivative with respect  to a parameter $S$,$S^{\alpha \beta}$ is the spin tensor, \begin{equation} M^{\mu \nu} =P^{\mu}U^{\nu}- P^{\nu}U^{\mu} \end{equation}, regarding $U^{\mu}$ is the unit tangent vector to the geodesic. \\
 Using the following identity on both equations (2.20) and (2.21)
\begin{equation}
  A^{\mu}_{; \nu \rho} - A^{\mu}_{; \rho \nu} = R^{\mu}_{\beta \nu \rho} A^{\beta},
 \end{equation}
  where $A^{\mu}$ is an arbitrary vector, and multiplying both sides with arbitrary vectors, $U^{\rho} \Psi^{\nu}$ as well as using the following condition [26]
\begin{equation}
 U^{\alpha}_{; \rho} \Psi^{\rho} =  \Psi^{\alpha}_{; \rho } U^{\rho},
 \end{equation}
and $\Psi^{\alpha}$ is its deviation vector associated to the  unit vector tangent $U^{\alpha}$.
 Also in a similar way:
\begin{equation}
 S^{\alpha \beta}_{; \rho} \Psi^{\rho} =  \Phi^{\alpha \beta}_{; \rho } U^{\rho}
 \end{equation}

 One obtains the corresponding deviation equations [27]
\begin{equation}
\frac{D^2 \Psi^{\mu}}{DS^2}= R^{\mu}_{\nu \rho \sigma}P^{\nu} U^{\rho} \Psi^{\sigma}+ F^{\mu}_{; \rho} \Psi^{\rho},
  \end{equation}
and
\begin{equation}
\frac{D^2\Psi^{\mu \nu}}{DS^2}=  S^{\rho [ \mu} R^{\nu ]}_{\rho \sigma \epsilon} U^{\sigma} \Psi^{\epsilon} + M^{\mu \nu}_{; \rho} \Psi^{\rho} \end{equation}
\section{Motion in AP-Geometry}
\subsection*{ A brief introduction of AP-Space}
The structure of this space is defined completely by a set of n-contravariant vector fields $\h{i}^{\mu}$ where
$i =1,2,3,...,n)$ denotes the vector number, and $\mu ( = 1,2,3,...,n)$
denotes
$\h{i}_{\mu}$ of the vectors $\h{i}^{\mu}$, in the determinant  $||
\h{i}^{\mu} ||$, is defined such that\footnote{ for more detail see [13-16]}
$$
 \h{i}^{\mu}\h{j}_{\mu} = \delta_{ij},
$$
$$
\h{i}^{\mu}\h{i}_{\nu} = \delta^{\mu}_{\nu}.
$$
Using these vectors, the following second order symmetric tensors
are defined:
$$
 g^{\mu \nu} \edf \h{i}^{\mu} \h{i}^{\nu} ,
 $$

$$
g_{\mu \nu} \edf \h{i}_{\mu} \h{i}_{\nu} .
$$
 one can define Christoffel symbols and covariant derivatives using this
symbol, in the usual manner. The following third order tensor, the
contortion tensor, can be defined as,
$$
\gamma^{\alpha}_{. \mu \nu} \edf \h{i}^{\alpha} \h{i}_{\mu ; \nu},
$$

which is non-symmetric in its last two indices $\mu , \nu$. It can
be shown that $\gamma^{\alpha}_{. \mu \nu}$ is skew-symmetric in
its first two indices.\\
{\bf{The AP-Condition}} \\
$$
\h{i}^{\nu}_{| \stackrel{\mu}{+}} =0
$$
where $| \stackrel{\mu}{+}$ is the absolute +ve derivative,
such that it defines $\Gamma^{\alpha}-{\mu \nu}$ a non symmetric affine connection, in which

$$
\Gamma^{\alpha}_{\mu \nu}= \h{i}^{\alpha} \h{i}_{\mu ,\nu}.
$$
\subsection*{{(i)}Paths and Path Deviation Equations Subject to  (+)ve Derivative}
Paths and Path deviations equations are the counterpart of geodesic and geodesic deviation in AP-geometry.
Accordingly, we have different trajectories based on the type of the absolute derivative [17].

From this perspective, it has been found out that the Bazanski Lagrangian may be a good candidate to express  these trajectories.
\begin{equation}
L = g_{\mu \nu} V^{\mu} \frac{\nabla \Phi^{\nu}}{\nabla S^{+}}
 \end{equation}
where $$\frac{\nabla \Phi}{\nabla S^{+}} = \frac{d \Phi^{\alpha}}{ d S^{+}} + \Gamma^{\alpha}_{\mu \nu} \Phi^{\mu} V^{\nu}.$$
Thus,taking the variation with respect to $\xi^{\mu}$ and implementing the AP-condition to find that
\begin{equation}
g_{\stackrel{\mu}{+} \stackrel{\sigma}{\nu}{+} | \sigma} \equiv 0
 \end{equation}
one finds out the following path equation
\begin{equation}
\frac{\nabla V^{\mu}}{\nabla S^{+}} = 0.
 \end{equation}

 Also, its associated  deviations can be derived if one applies  the following relation

\begin{equation}
A^{\mu}_{| \stackrel{\nu}{+}\stackrel{\rho}{+} } - A^{\mu}_{|\stackrel{\rho}{+}\stackrel{\nu}{+} } =  M^{\mu}_{\sigma \nu \rho} A^{\sigma}+ \Lambda ^{\sigma}_{\nu \rho} A^{\mu}_{|\stackrel{\sigma}{+} },
 \end{equation}
 provided with the following condition:
\begin{equation}
U^{\alpha}_{| \stackrel{\rho}{+}}\Phi^{\rho} = \Phi^{\alpha}_{| \stackrel{\rho}{+}}U^{\rho},
\end{equation}
 with taking into consideration,  the vanishing curvature tensor
\begin{equation}
M^{\mu}_{\sigma \nu \rho} \equiv 0,
 \end{equation}
to be substituted in (3.30)  to obtain the corresponding deviation equation
\begin{equation}
\frac{\nabla^{2} \Phi^{\alpha}}{\nabla^{2+}} =  \Lambda^{\rho}_{\mu \nu} V^{\mu} \Phi^{\nu} V^{\alpha}_{| \stackrel {\rho}{+}}.
 \end{equation}
\subsection*{{(ii)}Paths and Path Deviation Equations subject to  (0)ve Derivative}
Due to (0)ve derivative one can derive its associated  path and path deviation equations using the following Lagrangian [17]:
\begin{equation}
L = g_{\mu \nu} W^{\mu} \frac{\hat{\nabla} \eta^{\nu}}{\hat{\nabla}S^{0}},
\end{equation}
where $$\frac{\hat{\nabla} \eta^{\alpha}}{\hat{\nabla} S^{0}} = \frac{d \eta^{\alpha}}{ d S^{0}} + \Gamma^{\alpha}_{(\mu \nu)} \eta^{\mu} W^{\nu} .$$

Thus, taking the variation with respect to $\eta^{\mu}$;provided that
\begin{equation}
g_{\stackrel{\mu}{0}\stackrel{\nu}{0}| \sigma } =  \Lambda_{(\mu \nu) \sigma},
 \end{equation}

to obtain its corresponding  path equation:
\begin{equation}
\frac{\nabla W^{\mu}}{\nabla S^{0}} =   \frac{1}{2} \Lambda^{~~ \mu}_{(\nu \rho).} W^{\nu}W^{\rho}.
 \end{equation}

Using the following relation

\begin{equation}
A^{\mu}_{| \stackrel{\nu}{0}\stackrel{\rho}{0} } - A^{\mu}_{|\stackrel{\rho}{0}\stackrel{\nu}{0} } = L^{\mu}_{\sigma \nu \rho} A^{\sigma}  + \Lambda ^{\sigma}_{\nu \rho} A^{\mu}_{|\stackrel{\sigma}{0} },
 \end{equation}
and  the condition below
\begin{equation}
W^{\mu}_{\stackrel{\rho}{(0)}} \eta^{\rho} = \eta^{\mu}_{\stackrel{\rho}{(0)}} W^{\rho},
\end{equation}
to be substituted in (3.37), provided that its associated curvature,

\begin{equation}
L^{\mu}_{\sigma \nu \rho} \neq 0,
\end{equation}
 is non vanishing, to obtain the  corresponding deviation equation
\begin{equation}
\frac{\hat{\nabla}^{2} \zeta^{\alpha}}{\hat{\nabla} S^{2(0)}} = \frac{1}{2} (\Lambda^{.~.~ \alpha}_{\mu \nu}W^{\mu}W^{\nu})_{| \stackrel{\rho}{0}}\zeta^{\rho}+   L^{\alpha}_{\beta \rho \sigma} W^{\beta} W^{\rho} \zeta^{\sigma} + \Lambda^{\rho}_{\mu \nu} W^{\mu} \eta^{\nu}\zeta^{\alpha}_{| \stackrel{\rho}{0}}
 \end{equation}

\subsection*{{(iii)}Paths and Path Deviation Equations Subject to  (-)ve Derivative}
 Following the same approach as explained the previous items(i) and (ii), one may derive the paths and path deviations equations associated to -ve derivative, by introducing the following Lagrangian [17]:
\begin{equation}
L = g_{\mu \nu} J^{\mu} \frac{\tilde{\nabla} \zeta^{\nu}}{\tilde{\nabla}S^{-}}
 \end{equation}
 such that
  $$ \frac{\tilde{\nabla} \zeta^{\nu}}{\tilde{\nabla}S^{-}} =  \frac{d \zeta^{\nu}}{d S^{-}}+ \tilde{\Gamma}^{\nu}_{ \mu \sigma} \zeta^{\mu} J^{\sigma}.$$
Accordingly,  taking the variation with respect to $\eta^{\mu} $ to derive its corresponding path equation, and provided that [16]
\begin{equation}
g_{\stackrel{\mu}{-} \stackrel{\nu}{-} | \sigma} =  2\Lambda_{(\mu \nu) \sigma}
 \end{equation}
 to get
\begin{equation}
\frac{\tilde{\nabla} J^{\mu}}{\tilde{\nabla}S^{-}} =  \Lambda_{(\alpha \beta)}^{~.~ \mu} J^{\alpha} J^{\beta} .
 \end{equation}
Also, in order to derive its corresponding path deviation equation, one must take into account the following relation:
\begin{equation}
A^{\mu}_{| \stackrel{\nu}{-}\stackrel{\rho}{-} } - A^{\mu}_{| \stackrel{\rho}{-}\stackrel{\nu}{-} } = N^{\mu}_{\sigma \nu \rho} A^{\sigma}  + \Lambda ^{\sigma}_{\nu \rho} A^{\mu}_{|\stackrel{\sigma}{-} }
 \end{equation}
together with, the  condition
\begin{equation}
J^{\mu}_{| \stackrel{\rho}{(-)}} \zeta^{\rho} = \zeta^{\mu}_{| \stackrel{\rho}{(-)}} J^{\rho},
\end{equation}
to be substituted in (3.44), provided that its associated curvature,
\begin{equation}
N^{\mu}_{\sigma \nu \rho} \neq 0,
 \end{equation}
is non vanishing curvature.

 Thus we derive the corresponding path deviation equation
\begin{equation}
\frac{\tilde{\nabla}^{2} \eta^{\alpha}}{\tilde{\nabla}S^{2-}} =  N^{\alpha}_{\beta \rho \sigma} J^{\beta} J^{\rho} \eta^{\sigma} + \Lambda^{\rho}_{\mu \nu} J^{\mu} \eta^{\nu}\eta^{\alpha}_{| \stackrel{\rho}{-}}
\end{equation}

\subsection{On the Relation Between Spin Tensor and The Deviation Vector: The AP-geometry}
In this part, we are going to extend the relationship obtained in (2.2) to derive the corresponding spin equations and their corresponding spin deviation equations.
\subsection*{+ve Derivative}
Equations of spinning motion ,the case of $P_{+}^{\alpha}=mV^{\alpha}$ can be related to geodesic if one follows the following transformation
\begin{equation}
\bar{V}^{\mu} = V^{\mu} + \beta \frac{D \Phi^{\mu}}{Ds^{+}}
 \end{equation}
where $\bar{V}^{\alpha}$ is  a unit tangent vector with respect to the parameter ,such that  $\bar{V}^{\alpha} = \frac{d x^{\mu}}{d \bar{s}^{+}} $ , $\bar{s}$.
By taking the covariant derivative on both sides one obtains:
\begin{equation}
\frac{ \nabla \bar{V}^{\alpha}}{\nabla \bar{s}^{+}}= \frac{\nabla}{\nabla s^{+}}(V^{\mu} + \beta \frac{\nabla \Phi^{\mu}}{\nabla {s}^{+}})\frac{ds}{d \bar{s}}.
 \end{equation}
Substituting equations (3.30) and (3.34) in  (3.50) to get
\begin{equation}
\frac{\nabla \bar{V}^{\alpha}}{\nabla \bar{s}^{+}}=( \beta \Lambda^{\rho}_{\nu \sigma} V_{\mu | \stackrel{\rho}{+} } V^{\nu} \Phi^{\sigma}
)\frac{ds}{d \bar{s}}
 \end{equation}
let us assume the following
Taking $\beta =  \frac{s}{m}$
\begin{equation}
\bar{S}^{\mu \nu} = s {(V^{\alpha }\Phi^{\beta}-V^{\beta }\Phi^{\alpha}) }
\end{equation}
Thus, we get
\begin{equation}
\frac{\nabla \bar{V}^{\alpha}}{\nabla \bar{s}^{+}}=\frac{1}{2m}( \Lambda^{\rho}_{\nu \sigma} V_{\mu | \stackrel{\rho}{+}} \bar{S}^{\nu \sigma} ) \frac{ds}{d \bar{s}}
 \end{equation}

i.e
\begin{equation}
\frac{\nabla \bar{V}^{\alpha}}{ \nabla \bar{s}^{+}}= \frac{1}{2m} \Lambda^{\rho}_{\nu \sigma} \bar{V}_{\mu | \stackrel{\rho}{+}} \bar{S}^{\nu \sigma}
 \end{equation}

Which is the version the Papapetrou equation for +ve derivative, for short.
\subsection*{0-Derivative}
Equations of spinning motion ,the case of $P_{(0)}^{\alpha}=mW^{\alpha}$ can be related to geodesic if one follows the following transformation
\begin{equation}
\bar{W}^{\mu} = W^{\mu} + \beta \frac{\bar{\nabla} \eta^{\mu}}{\bar{\nabla}\bar{s}^{(0)}}
\end{equation}
where $\bar{W}^{\alpha}$ is  a unit tangent vector with respect to the parameter ,such that  $\bar{W}^{\alpha} = \frac{d x^{\mu}}{d \bar{s}^{(0)}} $ , $\bar{s}$.
By taking the covariant derivative on both sides one obtains:
\begin{equation}
\frac{\bar{\nabla} \bar{W}^{\alpha}}{\bar{\nabla} \bar{s}^{(0)}}= \frac{\nabla}{\nabla s^{+}}(W^{\mu} + \beta \frac{\nabla \eta^{\mu}}{\nabla {s}^{(0)}})\frac{ds}{d \bar{s}}.
 \end{equation}
Substituting equations (3.37) and (3.41) in  (3.56) to get
\begin{equation}
\frac{\nabla \bar{W}^{\alpha}}{\nabla \bar{s}^{(0)}}=( \frac{1}{2}  \Lambda^{~.~. \alpha }_{\mu \nu} W^{\mu}W^{\nu} + \beta [ L^{\alpha}_{ \beta \gamma \delta} W^{\beta}W^{\gamma}\eta^{\delta}  + \Lambda^{\rho}_{\nu \sigma} W_{\mu | \stackrel{\rho}{(0)} } W^{\nu} \eta^{\sigma} ]
)\frac{ds}{d \bar{s}}.
 \end{equation}
Now, let us assume that $\beta =  \frac{s}{m}$, and
\begin{equation}
\bar{S}^{\mu \nu} = s {(V^{\alpha }\eta^{\beta}-W^{\beta }\eta^{\alpha}) }.
\end{equation}
Thus, we get
\begin{equation}
\frac{\bar{\nabla} \bar{W}^{\alpha}}{\nabla \bar{s}^{(0)}}=( \frac{1}{2} \Lambda^{~.~. \alpha }_{\mu \nu} W^{\mu}W^{\nu} + \frac{1}{2m}[ L^{\alpha}_{\mu \nu \sigma} W^{\mu} + \Lambda^{\rho}_{\nu \sigma} W_{\mu | \stackrel{\rho}{(0)}}] ) \bar{S}^{\nu \sigma} ( \frac{ds^{(0)}}{d \bar{s}^{(0)}})
 \end{equation}

i.e
\begin{equation}
\frac{\nabla \bar{W}^{\alpha}}{ \nabla \bar{s}^{(0)}}= \frac{1}{2} \Lambda^{~.~. \alpha }_{\mu \nu} W^{\mu}\bar{W}^{\nu} + \frac{1}{2m} (L^{\alpha}_{\mu \nu \sigma} \bar{W}^{\mu} + \Lambda^{\rho}_{\nu \sigma} \bar{W}_{\mu | \stackrel{\rho}{(0)}}) \bar{S}^{\nu \sigma}.
 \end{equation}

If we regard
 $$
\frac{ds^{(0)}}{d \bar{s}^{(0)}} =1,
$$
then, equation (3.64) becomes
\begin{equation}
\frac{\nabla \bar{W}^{\alpha}}{ \nabla \bar{s}^{(0)}}= \frac{1}{2} \Lambda^{~.~. \alpha }_{\mu \nu} \bar{W}^{\mu}\bar{W}^{\nu} + \frac{1}{2m} (L^{\alpha}_{\mu \nu \sigma} \bar{W}^{\mu} + \Lambda^{\rho}_{\nu \sigma} \bar{W}_{\mu | \stackrel{\rho}{(0)}}) \bar{S}^{\nu \sigma}
 \end{equation}

Which is the version the Papapetrou equation for (0)ve derivative, for short.

\subsection*{-ve Derivative}
Equations of spinning motion ,the case of $P_{-}^{\alpha}=mJ^{\alpha}$ can be related to geodesic if one follows the following transformation

\begin{equation}
\bar{J}^{\mu} = J^{\mu} + \beta \frac{\tilde{\nabla} \zeta^{\mu}}{\tilde{\nabla}\bar{s}^{-}}
 \end{equation}
where $\bar{J}^{\alpha}$ is  a unit tangent vector with respect to the parameter ,such that  $\bar{J}^{\alpha} = \frac{d x^{\mu}}{d \bar{s}^{-}} $ , $\bar{s}$.
By taking the covariant derivative on both sides one obtains:
\begin{equation}
\frac{\bar{\nabla} \bar{J}^{\alpha}}{\bar{\nabla} \tilde{s}^{(-)}}= \frac{\tilde{\nabla}}{\nabla s^{-}}(J^{\mu} + \beta \frac{\nabla \zeta^{\mu}}{\tilde{\nabla} {s}^{-}})\frac{ds}{d \bar{s}^{-}}.
 \end{equation}
Substituting equations (3.44) and (3.48) in  (3.63) to get
\begin{equation}
\frac{\tilde{\nabla} \bar{J}^{\alpha}}{\tilde{\nabla} \tilde{s}^{-}}=( \Lambda^{~.~. \alpha }_{\mu \nu} J^{\mu}J^{\nu}+ \beta [ N^{\alpha}_{\beta \gamma \delta} J^{\beta} J^{\gamma} \zeta^{\delta} + \Lambda^{\rho}_{\nu \sigma} J_{\mu | \stackrel{\rho}{-} } J^{\nu} \zeta^{\sigma}]
)\frac{ds^{-}}{d \bar{s^{-}}}
 \end{equation}
let us assume the following
Taking $\beta =  \frac{s}{m}$,and
\begin{equation}
\tilde{S}^{\mu \nu} = s {(J^{\alpha }\zeta^{\beta}-J^{\beta }\zeta^{\alpha}) }.
 \end{equation}
Thus, we get
\begin{equation}
\frac{\tilde{\nabla} \bar{J}^{\alpha}}{\tilde{\nabla} \bar{s}^{-}}=( \Lambda^{~.~. \alpha }_{\mu \nu} J^{\mu}J^{\nu}+ \frac{1}{2m} [ N^{\alpha}_{\mu \nu \sigma} J^{\mu} + \Lambda^{\rho}_{\nu \sigma} J_{\mu | \stackrel{\rho}{-}}]) \tilde{S}^{\nu \sigma} ( \frac{ds^{(-)}}{d \bar{s}^{(-)}})
 \end{equation}

i.e
\begin{equation}
\frac{\tilde{\nabla} \bar{J}^{\alpha}}{ \tilde{\nabla} \bar{s}^{-}}= \Lambda^{~.~. \alpha }_{\mu \nu} J^{\mu}\tilde{J}^{\nu} + \frac{1}{2m} ( N^{\alpha}_{\mu \nu \sigma} \bar{J}^{\mu} + \Lambda^{\rho}_{\nu \sigma} \bar{J}_{\mu | \stackrel{\rho}{-}}) \tilde{S}^{\nu \sigma}
 \end{equation}
 If we regard
 $$
\frac{ds^{(-)}}{d \tilde{s}^{(-)}} =1 ,
$$
then, equation (3.73) becomes
\begin{equation}
\frac{\tilde{\nabla} \bar{J}^{\alpha}}{ \tilde{\nabla} \bar{s}^{-}}= \Lambda^{~.~. \alpha }_{(\mu \nu)} \bar{J}^{\mu} \bar{J}^{\nu} + \frac{1}{2m} ( N^{\alpha}_{\mu \nu \sigma} \bar{J}^{\mu} + \Lambda^{\rho}_{\nu \sigma} \bar{J}_{\mu | \stackrel{\rho}{-}}) \tilde{S}^{\nu \sigma}
 \end{equation}
Which is the version the Papapetrou equation for -ve derivative, for short.
 \section{Spinning and Spinning Deviation Equations in AP-geometry: Lagrangian Formalism}
From the previous results, we can check the reliability of the corresponding Bazanski equation to become in the following way.
\subsection{Spinning and Spinning Deviation equations subject to (+)ve derivative }
\subsection*{{i} the case of $P_{+} = mV$}

 \begin{equation}
   L= g_{\mu \nu} V^{\mu}\frac{\bar{\nabla} \Phi ^{\mu}}{\bar{\nabla}S^{+}} + \bar{S}_{\mu \nu} \frac{\bar{\nabla} \Phi^{\mu \nu}}{ \bar{\nabla}S^{+}} + \frac{1}{2m} [\Lambda^{\rho}_{\nu \sigma} {V}_{\mu | \stackrel{\rho}{+}}] \bar{S}^{\nu \sigma} \Phi^{\mu}.
 \end{equation}
 Taking the variation with respect to $\Phi^{\alpha}$ and $\Phi^{\alpha \beta}$ we obtain
 \begin{equation}
 \frac{{\nabla} V^{\alpha}}{{\nabla}S^{+}} = \frac{1}{2m}\Lambda_{\rho \delta \nu}\bar{S}^{\delta \nu}V^{\alpha}_{\stackrel{\sigma}{+}} + g^{\alpha \rho} \Lambda^{\rho}_{\nu \sigma} {V}_{\mu | \stackrel{\rho}{+}} \bar{S}^{\nu \sigma},
 \end{equation}
    and

   \begin{equation}
    \frac{\tilde{\nabla} S^{\alpha \beta}}{\tilde{\nabla}S^{2-}} =  0
 \end{equation}
 Using the commutation relation (3.31) ,  conditions (3.32)  and

   \begin{equation}
    \bar{S}^{\mu \nu}_{| \stackrel{\rho}{+}} \Phi^{\rho} = \Phi^{\mu \nu}_{| \stackrel{\rho}{+}} V^{\rho},
   \end{equation}
  to be substituted in (4.70) and (4.71) in order to derive its corresponding set of deviation equations
\begin{equation}
 \frac{{\nabla}^{2} \Phi^{\alpha}}{\tilde{\nabla}S^{2(+)}} =  \Lambda^{\rho}_{\mu \nu}V^{\mu} V^{\nu} \Phi^{\alpha}_{\stackrel{\rho}{+}} ,
 \end{equation}

 and
 \begin{equation}
  \frac{{\nabla}^{2} \Phi^{\alpha \beta}}{\hat{\nabla}S^{2+}} = \Lambda^{\rho}_{\mu \nu}V^{\mu} \Phi^{\nu} \bar{S}^{\alpha \beta}_{| \stackrel{\rho}{+}}.
 \end{equation}

\subsection*{ (ii) the case $ P_{+} \neq m V$}
Let us suggest the following Lagrangian:
  \begin{equation}
  L= g_{\mu \nu} P_{+}^{\mu}\frac{{\nabla} \Phi ^{\mu}}{{\nabla}S^{+}} + \bar{S}_{\mu \nu} \frac{{\nabla}{\Phi}^{\mu \nu}}{{\nabla}S^{+}} + \frac{1}{2m} g_{\mu \nu} \Lambda^{\rho}_{\delta \rho}\bar{S}^{\delta \rho}V^{\mu}_{\stackrel{\rho}{+}}\Phi^{\nu} + g_{\alpha \mu }g_{\beta \nu}[ P_{+}^{\alpha}V^{\beta} -P_{+}^{\beta}V^{\alpha} ] \Phi^{\mu \nu},
 \end{equation}
where $$P_{+}^{\mu}= m V^{\mu}+ \frac{V_{\nu} \tilde{\nabla} \bar{S}^{\mu \nu}}{\nabla S^{(+)}}.$$
Taking the variation with respect to $\zeta^{\alpha}$ and $\zeta^{\alpha \beta}$ we obtain
\begin{equation}
 \frac{ {\nabla} P_{+}^{\alpha}}{{S}^{+}} = \frac{1}{2m} \Lambda^{\rho}_{ \delta \nu}\bar{S}^{\delta \nu}V^{\alpha}_{\stackrel{\rho}{+}} + g_{\mu \rho}g_{\nu \delta} [P_{+}^{\rho} V^{\delta}-P_{+}^{\delta} V^{\rho}  ] \Phi^{\mu \nu},
 \end{equation}
    and
\begin{equation}
    \frac{{\nabla}^{2} \bar{S}^{\alpha \beta}}{\bar{\nabla}{S}^{2}} =  [P_{+}^{\alpha}V^{\beta} -P_{+}^{\beta}V^{\alpha} ].
 \end{equation}
  Using the commutation relation (3.31) , the conditions (3.32) and (3.32) to be substituted in (4.70)and (4.71)and (4.72) in order to derive its corresponding set of deviation equations
\begin{equation}
 \frac{\tilde{\nabla}^{2} \Phi^{\alpha}}{\tilde{\nabla}S^{2-}} =  (  \frac{1}{2m}\Lambda^{\rho}_ {\delta \sigma}\hat{S}^{\delta \sigma}V^{\alpha}_{\stackrel{\sigma}{+}})_{| \stackrel{\delta}{+}} \Phi^{\delta}
 \end{equation}

 and
\begin{equation}
 \frac{{\nabla}^{2} \zeta^{\alpha \beta}}{\tilde{\nabla}S^{2+}} = \Lambda^{\rho}_{\mu \nu}V^{\mu} \Phi^{\nu} \tilde{S}^{\alpha \beta}_{\stackrel{\rho}{+}} +   [P_{+}^{\alpha}V^{\beta} -P_{+}^{\beta}V^{\alpha} ]_{| \stackrel{\delta}{+}}\Phi^{\delta}.
\end{equation}
 From the above results of spinning equations and their corresponding deviation ones, we reach to regard them as the equivalent set of equations of spinning objects in the presence of Tele-parallel gravity [13].
\subsection{Spinning and Spinning Deviation equations subject to (0)ve derivative }
{\subsection*{{ii} The case of ${P_{(0)} = mW}$}}

\begin{equation}
  L= g_{\mu \nu} W^{\mu}\frac{\hat{\nabla} \eta ^{\mu}}{\hat{\nabla}S^{(0)}} + \hat{S}_{\mu \nu} \frac{\hat{\nabla} \eta^{\mu \nu}}{ \hat{\nabla}S^{(0)}} + \frac{1}{2m} L_{\mu \nu \rho \delta} \eta^{\mu} W^{\nu} S^{\rho \delta}+ \Lambda^{\rho}_{\nu \sigma} {W}_{\mu | . \stackrel{\rho}{(0)}} \hat{S}^{\nu \sigma} \eta^{\mu}
  \end{equation}
 Taking the variation with respect to $\eta^{\alpha}$ and $\eta^{\alpha \beta}$, we obtain
\begin{equation}
 \frac{\hat{\nabla} W^{\alpha}}{\hat{\nabla}S^{o}} = \frac{1}{2} \Lambda^{~.~.~\alpha}_{(\mu \nu)} W^{\mu} W^{\nu} + \frac{1}{2m} L^{\alpha}_{\nu \rho \sigma} W^{\nu} S^{\rho \sigma} + g^{\alpha \mu} \Lambda^{\rho}_{\nu \sigma} {W}_{\mu | \stackrel{\rho}{(0)}} \hat{S}^{\nu \sigma}
 \end{equation}
    and

   \begin{equation}
    \frac{\hat{\nabla}^{2} \hat{S}^{\alpha \beta}}{\hat{\nabla}S^{2(0)}} = \frac{1}{2}\Lambda^{~.~.~[ \alpha }_{(\mu \nu)}S^{\beta ]\mu} W^{\nu}.
  \end{equation}
  Using the commutation relation (3.38) and  conditions (3.39) and

   \begin{equation}
    \hat{S}^{\mu \nu}_{| \stackrel{\rho}{(0)}} \eta^{\rho} = \eta^{\mu \nu}_{| \stackrel{\rho}{(0)}} W^{\rho}
   \end{equation}
  to be substituted in (4.80) and (4.81) in order to derive its corresponding set of deviation equations
\begin{equation}
 \frac{\hat{\nabla}^{2} \eta^{\alpha}}{\hat{\nabla}S^{2(0)}} = L^{\alpha}_{\mu \nu \rho} W^{\mu}W^{\nu}\eta^{\rho} + \Lambda^{\rho}_{\mu \nu}W^{\mu} \xi^{\nu} \eta^{\alpha}_{\stackrel{\rho}{(0)}} + \frac{1}{2} ( \Lambda^{~.~.~\alpha}_{\mu \nu} W^{\mu} W^{\nu} + \frac{1}{2m} L^{\alpha}_{\nu \rho \sigma} W^{\nu} S^{\rho \sigma})_{| \stackrel{\sigma}{0}}\eta^{\sigma},
 \end{equation}

 and
 \begin{equation}
 \frac{\hat{\nabla}^{2} \eta^{\alpha \beta}}{\hat{\nabla}S^{2(0)}} = S^{\mu [ \beta}{\rho}L^{\alpha ] }_{\mu \nu \rho} W^{\nu}\eta^{\rho} + \Lambda^{\rho}_{\mu \nu}W^{\mu} \eta^{\nu} S^{\alpha \beta}_{| \stackrel{\rho}{0}}.
 \end{equation}
\subsection*{ (ii) the case $ P_{(0)} \neq m W$}
Let us suggest the following Lagrangian:
  \begin{equation}
  L= g_{\mu \nu} P_{(0)}^{\mu}\frac{\hat{\nabla} \eta ^{\mu}}{\hat{\nabla}S^{(0)}} + \hat{S}_{\mu \nu} \frac{\hat{\nabla}{\eta}^{\mu \nu}}{\hat{\nabla}S^{(0)}} + \frac{1}{2m} L_{\mu \nu \rho \delta} \eta^{\mu} W^{\nu} \hat{S}^{\rho \delta}+ \frac{1}{2m} g_{\mu \nu} \Lambda^{\rho}_{\delta \rho}\hat{S}^{\delta \rho}W^{\mu}_{\stackrel{\rho}{(0)}}\eta^{\nu}+ g_{\mu \rho}g_{\nu \delta} [P_{(0)}^{\rho} W^{\delta}-P_{0}^{\delta} W^{\rho}  ] \eta^{\mu \nu}
 \end{equation}
where $P_{(0)}^{\mu}= m W^{\mu}+ \frac{W_{\nu} D S_{(0)}^{\mu \nu}}{D S^{(0)}}.$\\

 Taking the variation with respect to $\eta^{\alpha}$ and $\eta^{\alpha \beta}$ we obtain
\begin{equation}
 \frac{ \hat{\nabla} P_{(0)}^{\alpha}}{\hat{S}^{(0)}} = \frac{1}{2} \Lambda^{..\alpha}_{\mu \nu} P_{(0)}^{\mu} W^{\nu} + \frac{1}{2m} L^{\alpha}_{\nu \rho \sigma} W^{\nu} \hat{S}^{\rho \sigma} + \frac{1}{2m} \Lambda^{\rho}_{ \delta \nu}\hat{S}^{\delta \nu}W^{\alpha}_{| \stackrel{\rho}{o}}
 \end{equation}
    and
\begin{equation}
    \frac{\hat{\nabla}^{2} \hat{S}^{\alpha \beta}}{\hat{\nabla}{S}^{2(0)}} = \frac{1}{2} \Lambda^{~.~.[\alpha}_{\mu \nu} S^{\mu \beta]} W^{\nu}
 \end{equation}
  Using the commutation  relation (3.38) , conditions (3.39) and (4.82) to be substituted in (4.82) and (4.83)  in order to derive its corresponding set of deviation equations
\begin{equation}
 \frac{\hat{\nabla}^{2} \eta^{\alpha}}{\hat{\nabla}S^{2(0)}} = L^{\alpha}_{\mu \nu \rho} P_{(0)}^{\mu}W^{\nu}\eta^{\rho} + \Lambda^{\rho}_{\mu \nu}P_{(0)}^{\mu} \eta^{\nu} \eta^{\alpha}_{\stackrel{\rho}{(0)}} + \frac{1}{2} ( \Lambda^{..\alpha}_{\mu \nu} P_{(0)}^{\mu} W^{\nu} + \frac{1}{2m} L^{\alpha}_{\nu \rho \sigma} W^{\nu} \hat{S}^{\rho \sigma}+ \frac{1}{2m}\Lambda^{\rho}_ {\delta \sigma}\hat{S}^{\delta \sigma}W^{\alpha}_{\stackrel{\sigma}{(0)}})_{\stackrel{\delta}{(0)}} \eta^{\delta},
 \end{equation}

 and
\begin{equation}
 \frac{\tilde{\nabla}^{2} \zeta^{\alpha \beta}}{\tilde{\nabla}S^{2-}} = \tilde{S}^{\mu [ \beta}{\rho}N^{\alpha ] }_{\mu \nu \rho} J^{\nu} J^{\rho} + \Lambda^{\rho}_{\mu \nu}W^{\mu} \eta^{\nu} \hat{S}^{\alpha \beta}_{\stackrel{\rho}{o}} +   ( \Lambda^{~.~.[\alpha}_{\mu \nu} S^{\mu \beta]} J_{\nu| \stackrel{\delta}{-}}\zeta^{\delta} + [P_{(0)}^{\alpha}W^{\beta} -P_{(0)}^{\beta}W^{\alpha} ]_{| \stackrel{\delta}{(0)}}\eta^{\delta}
 \end{equation}
\subsection{Spinning and Spinning Deviation Equations subject to (-)ve Derivative }

\subsection*{{(i)} the case of $P_{-} = mJ$}

 \begin{equation}
   L= g_{\mu \nu} J^{\mu}\frac{\tilde{\nabla} \zeta ^{\mu}}{\tilde{\nabla}S^{-}} + \tilde{S}_{\mu \nu} \frac{\tilde{\nabla} \zeta^{\mu \nu}}{ \tilde{\nabla}S^{-}} + \frac{1}{2m} N_{\mu \nu \rho \delta} \zeta^{\mu} J^{\nu} S^{\rho \delta}+ \Lambda^{\rho}_{\nu \sigma} {J}_{\mu | \stackrel{\rho}{-}} \hat{S}^{\nu \sigma} \zeta^{\mu}
 \end{equation}
 by taking the variation with respect to $\zeta^{\alpha}$ and $\zeta^{\alpha \beta}$ we obtain
 \begin{equation}
 \frac{\tilde{\nabla} J^{\alpha}}{\tilde{\nabla}S^{-}} = \Lambda^{..\alpha}_{\mu \nu} J^{\mu} J^{\nu} + \frac{1}{2m} N^{\alpha}_{\nu \rho \sigma} J^{\nu} S^{\rho \sigma}+
 \frac{1}{2m}\Lambda_{\rho \delta \nu}S^{\delta \nu}J^{\alpha}_{\stackrel{\sigma}{-}} + g^{\alpha \rho} \Lambda^{\rho}_{\nu \sigma} {J}_{\mu | \stackrel{\rho}{-}} \tilde{S}^{\nu \sigma}
 \end{equation}
    and

   \begin{equation}
    \frac{\tilde{\nabla} S^{\alpha \beta}}{\tilde{\nabla}S^{2-}} =  \Lambda^{~.~ [\alpha }_{(\mu \nu)}S^{\beta]\mu} J^{\nu}
 \end{equation}
 Using the commutation relation (3.44) ,  conditions (3.48) and   
   \begin{equation}
    \tilde{S}^{\mu \nu}_{| \stackrel{\rho}{-}} \zeta^{\rho} = \zeta^{\mu \nu}_{| \stackrel{\rho}{-}} J^{\rho}
   \end{equation}
to be substituted in (4.91)and (4.92) in order to derive its corresponding set of deviation equations

\begin{equation}
 \frac{\tilde{\nabla}^{2} \zeta^{\alpha}}{\tilde{\nabla}S^{2(-)}} = N^{\alpha}_{\mu \nu \rho} J^{\mu}J^{\nu}\zeta^{\rho} + \Lambda^{\rho}_{\mu \nu}J^{\mu} J^{\nu} \eta^{\alpha}_{\stackrel{\rho}{-}} +  ( \Lambda^{~.~.~\alpha}_{\mu \nu} J^{\mu} W^{\nu} + \frac{1}{2m} N^{\alpha}_{\nu \rho \sigma} J^{\nu} \tilde{S}^{\rho \sigma})_{| \stackrel{\sigma}{-}}\zeta^{\sigma}
 \end{equation}

 and
 \begin{equation}
 \frac{\tilde{\nabla}^{2} \zeta^{\alpha \beta}}{\hat{\nabla}S^{2(-)}} = S^{\mu [ \beta}{\rho}{\Lambda}^{\alpha ] }_{ \nu \rho} J^{\nu}\zeta^{\rho} + \Lambda^{\rho}_{\mu \nu}J^{\mu} \zeta^{\nu} \tilde{S}^{\alpha \beta}_{| \stackrel{\rho}{-}}.
 \end{equation}

\subsection*{ (ii) the case $ P_{-} \neq m J$}
Let us suggest the following Lagrangian:
  \begin{equation}
  L= g_{\mu \nu} P_{-}^{\mu}\frac{\tilde{\nabla} \zeta ^{\mu}}{\tilde{\nabla}S^{-}} + \tilde{S}_{\mu \nu} \frac{\tilde{\nabla}{\zeta}^{\mu \nu}}{\tilde{\nabla}S^{-}} + \frac{1}{2m} N_{\mu \nu \rho \delta} \zeta^{\mu} J^{\nu} \tilde{S}^{\rho \delta}+ \frac{1}{2m} g_{\mu \nu} \Lambda^{\rho}_{\delta \rho}\tilde{S}^{\delta \rho}J^{\mu}_{\stackrel{\rho}{-}}\zeta^{\nu} + g_{\mu \rho}g_{\nu \delta} [P_{-}^{\rho} J^{\delta}-P_{-}^{\delta} J^{\rho}  ] \zeta^{\mu \nu}
 \end{equation}
where $$P_{(0)}^{\mu}= m W^{\mu}+ \frac{J_{\nu} \tilde{\nabla} \tilde{S}^{\mu \nu}}{D S^{(0)}}$$

 by taking the variation with respect to $\zeta^{\alpha}$ and $\zeta^{\alpha \beta}$ we obtain
\begin{equation}
 \frac{ \tilde{\nabla} P_{-}^{\alpha}}{\tilde{S}^{-}} = \frac{1}{2} \Lambda^{~.~.~\alpha}_{\mu \nu} P_{-}^{\mu} J^{\nu} + \frac{1}{2m} N^{\alpha}_{\nu \rho \sigma} J^{\nu} \tilde{S}^{\rho \sigma} + \frac{1}{2m} \Lambda^{\rho}_{ \delta \nu}\tilde{S}^{\delta \nu}J^{\alpha}_{\stackrel{\rho}{-}}
 \end{equation}
    and
\begin{equation}
    \frac{\tilde{\nabla}^{2} \tilde{S}^{\alpha \beta}}{\hat{\nabla}{S}^{2-}} =  \Lambda^{~.~.[\alpha}_{\mu \nu} S^{\mu \beta]} J^{\nu}
 \end{equation}
  Using commutation relation (3.44) and the conditions (3.48) and (4.93) to be substituted in (4.97) and (4.98) in order to derive its corresponding set of deviation equations
\begin{equation}
 \frac{\tilde{\nabla}^{2} \zeta^{\alpha}}{\tilde{\nabla}S^{2-}} = N^{\alpha}_{\mu \nu \rho} P_{-}^{\mu}J^{\nu}\zeta^{\rho} + \Lambda^{\rho}_{\mu \nu}P_{-}^{\mu} \zeta^{\nu} \zeta^{\alpha}_{\stackrel{\rho}{o}} + ( \Lambda^{~.~.~\alpha}_{\mu \nu} P_{-}^{\mu} J^{\nu} + \frac{1}{2m} L^{\alpha}_{\nu \rho \sigma} J^{\nu} \tilde{S}^{\rho \sigma}+ \frac{1}{2m}\Lambda^{\rho}_ {\delta \sigma}\hat{S}^{\delta \sigma}J^{\alpha}_{\stackrel{\sigma}{-}})_{| \stackrel{\delta}{-}} \zeta^{\delta}
 \end{equation}

 and
\begin{equation}
 \frac{\tilde{\nabla}^{2} \zeta^{\alpha \beta}}{\tilde{\nabla}S^{2-}} = \tilde{S}^{\mu [ \beta}{\rho}N^{\alpha ] }_{\mu \nu \rho} J^{\nu} J^{\rho} + \Lambda^{\rho}_{\mu \nu}J^{\mu} \zeta^{\nu} \tilde{S}^{\alpha \beta}_{\stackrel{\rho}{-}} +   ( \Lambda^{~.~.[\alpha}_{\mu \nu} \tilde{S}^{\mu \beta]} J_{\nu| \stackrel{\delta}{-}}\zeta^{\delta}+ [P_{-}^{\alpha}J^{\beta} -P_{-}^{\beta}J^{\alpha} ]_{\stackrel{\delta}{-}}\zeta^{\delta}.
 \end{equation}
\section{Discussion and Concluding Remarks}
The present work is related to extending   the concept of geometerization of physics to explain spinning  objects in a gravitational field. It has been developed the modified Bazanski Lagrangian in general relativity for spinning objects to be expressed in AP-geometry. Due to the wealth of geometric quantities, one must regard that the existence of spin tensors associated for each path is  defined by a specific type of absolute derivative. Also, we have emphasized  the relationship between geodesic and geodesic deviation with spinning tensors, to be viable for any type of geometries,  by testing its reliability in both Riemannian and AP-geometry.
Moreover, the spin tensor has been defined geometrically as a commutation relation between formula between  geodesic and geodesic deviation in Riemannian geometry and their counterparts in AP-geometry.

Accordingly, we have obtained three different spinning equations different from its counterpart in Riemannian geometry.  One of them, can be used to describe the spinning equations and their deviation in Tele-parallel gravity, i.e. these sets of spinning equations are representing the Papapetrou equation of Hayshi-Shirifugi New General Relativity [13],  while the other two paths may describe, hypnotically, a set of spinning particles subject to a class of non vanishing curvature and torsion simultaneously. This may present require an efficient field theory feasible to give a physical interpretation of  $\hat{S}^{\mu \nu}$ and $\tilde{S}^{\mu \nu} $ which still an open question.\\

Yet, this study has also clarified the viability interaction between torsion tensor and spin deviation equations, as mentioned previously in case of Gauge theories of gravity [24]
Nevertheless, these sets of spinning equations can also be applied in PAP-geometry, to give new results. Owing to revisit, the bi-metric theories of gravity using the  tetrad formalism, one may find out some promising results able to reveal the mystery of several anomalies such as dark matter and dark energy in our nature, which will be studied in our future work.

\section*{References}
{[1]} M. Mathisson, Acta Phys. Polon {\bf{6}}, 163 (1937).\\
{[2]} A.Papapetrou , Proceedings of Royal Society London A {\bf{209}} , 248(1951).  \\
{[3]} W. G. Dixon   Proc. R. Soc. London, Ser. A {\bf{314}}, 499 (1970). \\
{[4]}  F. Cianfrani, I. Milillo and G. Montani Phys, Lett. A,  {\bf{366}},7 ; gr-qc/0701157 (2007).  \\
{[5]}Utyiama, R.(1956) Phys Rev. 101 1597 \\
{[6]}Kibble, T.W. (1960) J. Math Phys., 2, 212 \\
{[7]}Hehl, von der Heyde, P. Kerlik, G.D. and Nester, J.M. (1976) Rev Mod Phys 48, 393-416 \\
{[8]} F.W. Hehl , Proceedings of the 6th Course of the International School of Cosmology and Gravitation on "Spin, Torsion and Supergravity"  eds. P.G. Bergamann  and V. de Sabatta , held at Erice ,1 (1979).\\
{[9]}H.I. Acros, and J.G. Pereira, International Journal of Modern Physics D  {\bf{13}}, 2193 (2004) \\
{[10]} S. Hojman, Physical Rev. D  {\bf{18}}, 2741 (1978). \\
([11])P.H. Yasskin, and W.R. Stoeger, Phys. Rev. D 21, 2081(1980). \\
{[12]}Hammond, R. (2002) Rep. Prog. Phys, 65, 599-449 \\
{[13]}K. Hayashi,  and T. Shirifuji , Phys. Rev. D, {\bf{19}}, 3524 (1979). \\
{[14]} Mikhail, F.I. and Wanas, M.I. (1977) Proc. Roy. Soc. Lond.
{\bf{A 356}}, 471. \\
{[15]} Wanas, M.I. (2001) Stud. Cercet. \c
Stiin\c t. Ser. Mat. Univ. Bac\u au {\bf{10}}, 297;
 gr-qc/0209050 \\
 {[16]} Wanas, M.I. (2000) Turk. J. Phys., {\bf{24}}, 473 ;
gr-qc/0010099. \\
{[17]} Wanas, M.I., Melek, M. and Kahil, M.E.(1995) Astrophys. Space
Sci.,{\bf{228}}, 273. ; gr-qc/0207113. \\
 {[18]} Wanas, M.I.(1998) Astrophys. Space Sci.,
{\bf{258}}, 237 ;vgr-qc/9904019. \\ \\
 {[19]} Wanas, M.I., Melek, M. and Kahil, M.E. (2000) Grav.
Cosmol., {\bf{6} }, 319. \\ \\
{[20]} Wanas, M.I., Melek, M. and Kahil, M.E. (2002) Proc. MG IX,
part B, p.1100, Eds.
V.G. Gurzadyan et al. (World Scientific Pub.); gr-qc/0306086. \\ \\
{[21]} Bazanski, S.I. (1989) J. Math. Phys.,
{\bf{30}}, 1018.\\
{[22]} D. Bini and A Geralico, Phys. Rev D {\bf{{84}}},104012; arXiv: 1408.4952 (2011)\\
{[23]} Kahil, M.E. (2006)  , J. Math. Physics {\bf {47}},052501. \\
{[24]} Magd E. Kahil, Odessa Astronomical Publications, {\bf{vol {28/2}}}, 126. (2015) \\
{[25]} Magd E. Kahil ,   Gravi. Cosmol. {\bf{24}}, 83 (2018) \\
{[26]} M. Mohseni , Gen. Rel. Grav., {\bf{42}}, 2477 (2010). \\
{[27]}M. Roshan, Phys.Rev. D{\bf{87}},044005 ; arXiv 1210.3136 (2013)\\

\end{document}